\documentclass[sigconf]{aamas} 

\usepackage{balance} 

\setcopyright{ifaamas}
\acmConference[AAMAS '26]{Proc.\@ of the 25th International Conference
on Autonomous Agents and Multiagent Systems (AAMAS 2026)}{May 25 -- 29, 2026}
{Paphos, Cyprus}{C.~Amato, L.~Dennis, V.~Mascardi, J.~Thangarajah (eds.)}
\copyrightyear{2026}
\acmYear{2026}
\acmDOI{}
\acmPrice{}
\acmISBN{}

\usepackage{caption}
\usepackage{subcaption}
\usepackage{csquotes}
\MakeOuterQuote{"}
\usepackage{booktabs,tabularx,array} 
\usepackage{adjustbox}

\newcommand{\deleted}[1]{\textcolor{lightgray}{}}

\begin{document}

\title{Insured Agents: A Decentralized Trust Insurance Mechanism for Agentic Economy}

\author{Botao `Amber' Hu}\authornote{Corresponding author}
\orcid{0000-0002-4504-0941}
\affiliation{%
  \institution{University of Oxford}
  \city{Oxford}
  \country{UK}
  }
\email{botao.hu@cs.ox.ac.uk}

\author{Bangdao Chen}
\orcid{}
\affiliation{%
  \institution{University College Oxford Blockchain Research Center}
  \city{Oxford}
  \country{UK}
  }
\email{bangdao.chen@gmail.com}



\begin{abstract}
The emerging “agentic web” envisions large populations of autonomous agents coordinating, transacting, and delegating across open networks. Yet many agent communication and commerce protocols treat agents as low-cost identities, despite the empirical reality that LLM agents remain unreliable, hallucinated, manipulable, and vulnerable to prompt-injection and tool-abuse. A natural response is “agents-at-stake”: binding economically meaningful, slashable collateral to persistent identities and adjudicating misbehavior with verifiable evidence. However, heterogeneous tasks make universal verification brittle and centralization-prone, while traditional reputation struggles under rapid model drift and opaque internal states. We propose a protocol-native alternative: insured agents. Specialized insurer agents post stake on behalf of operational agents in exchange for premiums, and receive privileged, privacy-preserving audit access via TEEs to assess claims. A hierarchical insurer market calibrates stake through pricing, decentralizes verification via competitive underwriting, and yields incentive-compatible dispute resolution.
\end{abstract}

\maketitle

\section{Introduction}

The next phase of internet is likely an "agentic web"—an internet of autonomous agents that discover each other, negotiate tasks, invoke tools, and execute payments with minimal human intervention \cite{Yang2025Agenticd,Wang2025Internet}. Recent protocol and standardization efforts aim to support this vision, including agent-to-agent messaging and orchestration tools like the Model Context Protocol (MCP) \cite{Hou2025Model}, agent communication protocols such as A2A \cite{a2a}, payment-layer proposals like AP2 \cite{ap2} and x402 \cite{x402}, agentic validation registries \cite{erc8004}, and the broader trends toward an agentic economy \cite{Vaziry2025MultiAgent,Rothschild2025Agentic}. 

AI safety research has shown that LLM agents can be neither consistently reliable \cite{Cemri2025Why} nor robust to adversarial manipulation \cite{Zou2023Universal} and induced into unsafe actions through hallucination \cite{Xu2025Hallucination}, hypersensitivity to nudges \cite{Cherep2025LLM}, deceptive behavior via prompt triggers \cite{Hubinger2024Sleeper}, and prompt injection \cite{Liu2024Prompt} and even prompt infection in multi-agent systems (MAS) \cite{Lee2024Prompt}. In open networks, we cannot assume that any LLM agent will be reliably trustworthy \cite{Salloum2024Trustworthiness}. It is better to assume trustlessness and build in mechanisms for credible accountability \cite{Werbach2016Trustless,Raji2020Closing}.


Current inter-agent protocols operate under "nothing-at-stake" dynamics \cite{Li2017Securing}. A natural response is the "agent-at-stake" principle—requiring agents to maintain slashable collateral tied to persistent identities and verifiable evidence—which appears well-suited for agentic economies. This mirrors traditional trust mechanisms, such as escrow in e-commerce \cite{ONeil1986Escrow}, and slashing-style cryptoeconomic security in proof-of-stake systems, where misbehavior triggers stake loss \cite{He2023Dont}. Emerging agent standards like ERC 8004 already anticipate "agents-at-stake" and validation registries, including stake-secured re-execution \cite{erc8004}. Yet universal verification of agent behavior or "proof of misbehavior" remains hard: unlike deterministic system behavior in blockchain, agentic tasks are \textbf{heterogeneous}, domain-dependent, private, and often involve ambiguous human preferences. Verification can become centralized, costly, or vulnerable to collusion \cite{Alger2003Moral}. Reputation systems face additional strain from the fluidity of LLM-based agents—identity shifts, model updates, memory decay, prompt changes, tool access changes—which undermines the reputation assumption of stable behavior over time.

We propose \textit{Insured Agents}: a hierarchical, voluntary insurance mechanism. Instead of requiring every agent to self-bond, an agent purchases a policy from an insurer. The insurer posts stake on the agent's behalf, earns premiums, and faces slashing if misbehavior is proven. In exchange, the insurer receives privileged audit access to the agent's internal traces (e.g., signed logs inside a TEE), enabling privacy-preserving adjudication without global public disclosure. The result is a competitive underwriting market that: (i) calibrates stakes to task-specific risk through underwriting, (ii) supports privacy-preserving verification via conditional audit access (e.g., TEEs and cryptographic proofs), and (iii) decentralizes verification by distributing liability across hierarchical, competing insurer agents—avoiding reliance on a single trusted verifier. This protocol frames trust as a market rather than a static reputation score, creating incentives to invest in monitoring, evaluation, and dispute resolution capacity.

This framing is consistent with recent AAMAS Blue Sky calls to design safe systems even when constituent agents are unsafe \cite{Nezami2025Safety}, build decentralized LLM-based MAS with privacy-preserving monetization \cite{Ding2025Decentralized}, and use markets as internal coordination mechanisms under uncertainty \cite{Sudhir2025Marketbased}. Our contribution is to introduce insurance as an MAS mechanism for designing protocols for open agent economies, together with a minimal incentive analysis and a research agenda.

\section{Background}
\subsection{Traditional trust mechanism: escrow, reputation, slashing, and insurance}

Escrow and mediated payments reduce risk in online trade by reallocating settlement authority and enabling conditional release, which complements reputational feedback \cite{Resnick2000Reputation,Josang2007survey,Sabater2005Review,Pinyol2013Computationala}. 
Online feedback mechanisms can discipline opportunistic behavior but are vulnerable to strategic manipulation, collusion, and sybil attacks \cite{Dellarocas2003Digitization,Josang2007survey,Douceur2002Sybil}. The “cheap pseudonyms” problem highlights how low-cost identity resets undermine accountability in open systems \cite{Friedman2001Sociala}. 

In cryptoeconomic systems, proof-of-stake (PoS) consensus and related designs use bonded collateral and slashing to make protocol violations expensive, aiming to deter equivocation and other strategic attacks \cite{Li2017Securing}. The "nothing-at-stake" problem—when identities are cheap and misbehavior is not costly—reappears in many open-agent settings and suggests that stake can substitute for trust, at least for well-defined violations. Yet extending PoS-style proofs to agentic tasks is non-trivial: verification is heterogeneous and domain-specific.

Insurance markets exist to pool risk, transfer liability, and create incentives through premiums, deductibles, and exclusions \cite{Rothschild1976Equilibrium}. They also expose classic incentive pathologies: adverse selection (high-risk parties disproportionately seek coverage) and moral hazard (coverage reduces incentives for care) \cite{Akerlof1970Market}. Insurance institutions often co-evolve with auditing, compliance, and standardized reporting to manage information asymmetries. Yet traditional insurance remains institutional: verification becomes centralized, costly, and vulnerable to corruption, collusion and domain-expertise mismatch. This leads to time-consuming processes, privacy leakage, and surveillance capitalism. The agentic economy demands that insurance evolve from institution to autonomous protocol.

\subsection{Trustworthy Multi-Agent Systems}

MAS research has long leveraged economic incentives to induce trustworthy behaviour under strategic interaction, via mechanism design, reputation engineering, and protocol analysis \cite{Myerson1981Optimal,Shoham2008Multiagent}, reputation \cite{Sabater2004EVALUATING, Teacy2006TRAVOS,josang2002beta,Huynh2006integrated} and trust models that reward reliable agents with future interaction opportunities \cite{Ramchurn2004Trust}, and automated negotiation/protocol design where small rule changes can shift equilibria and introduce vulnerabilities \cite{Jennings2001Automated,Mohammad2025}. Complementing agent-level incentives, system-level approaches argue that safety can be achieved even with unreliable agents by embedding monitoring, governance, and fail-safes in the surrounding infrastructure \cite{Bellay2025}; meanwhile, contestability and structured dispute processes connect MAS to recourse and procedural fairness for automated decisions \cite{dignum2025contesting,Wachter2017Why}. Finally, market-based internal architectures motivate a general principle for open agent economies: markets coordinate heterogeneous components under uncertainty, with prices summarizing risk and value \cite{Wellman1993MarketOriented,Sudhir2025Marketbased}; building on this, we propose protocol-native, market-based insurance to make “economic trust” explicit—agents insure commitments and face priced consequences for breach.

\section{The Insured Agent Mechanism}
\label{sec:insured-agents}

We propose \emph{insured agents} as a design pattern rather than a fully specified protocol. The core shift is from each agent bonding its own stake to agents outsourcing bonding and verification to specialized insurers that internalize losses. Trust in the agentic web is thus \emph{priced and underwritten}: instead of requiring every agent to self-stake large collateral, agents purchase coverage from insurer agents that pool capital, specialize in monitoring, and accept slashable liability. 

\subsection{The Mechanism}

\begin{figure}[ht]
    \centering
    \includegraphics[width=1\linewidth]{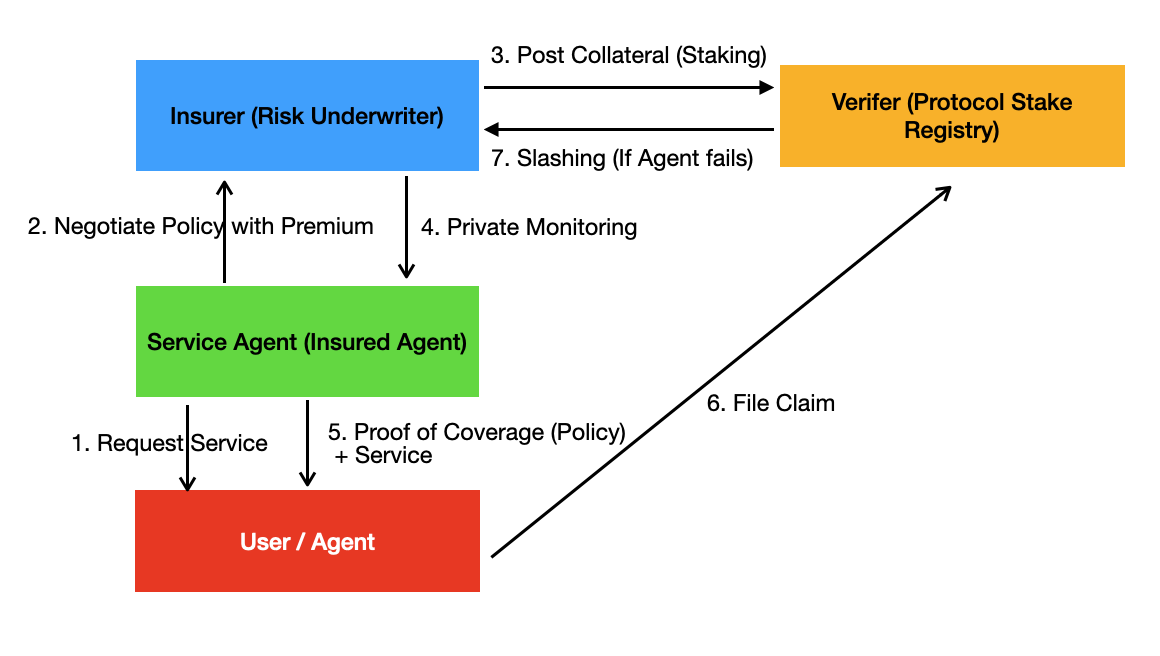}
    \caption{The Insured Agent Mechanism}
    \label{fig:insured}
\end{figure}

We consider four roles (implemented by agents or institutions):
\begin{itemize}
    \item \textbf{Service Agent} $A$: performs tasks (transactions, negotiations, delegated actions).
    \item \textbf{Insurer} $I$: underwrites $A$ by posting collateral and selling a policy.
    \item \textbf{User} $U$: delegates tasks to $A$ and may file claims.
    \item \textbf{Verifier} $V$: a decentralized arbitration layer that can adjudicate disputes at a cost.
\end{itemize}

A policy $\pi$ between $A$ and $I$ specifies: coverage amount, deductible, exclusions, admissible evidence, claim deadlines, challenge bonds, and the dispute-resolution route. To operate in high-stakes contexts, $A$ must present a valid policy. The insurer $I$ posts a slashable stake $S_I$ into a protocol registry linked to $A$’s identity; $A$ pays a premium $P$ and may post a deductible $S_A$ that $I$ can seize upon verified misbehavior. 

The high-level interaction is:
\begin{enumerate}
    \item \textbf{Underwriting.} $A$ obtains coverage from one or more insurers $I_1,\dots,I_n$, each posting stake $S_{I_k}$ and issuing a signed coverage credential. Policies may be compositional (base conduct policy plus domain-specific riders).
    \item \textbf{Proof of coverage.} When $A$ offers a service, it presents proofs of active coverage; users can require minimum coverage and specific insurers.
    \item \textbf{Claims and escalation.} If $U$ alleges harm, $U$ files a claim with evidence and a small bond. $I$ either settles immediately or denies, in which case $U$ may escalate by posting an additional bond, forcing $I$ to match and triggering $V$.
    \item \textbf{Adjudication and slashing.} $V$ evaluates evidence and decides. If the claim is valid, $U$ is compensated from $S_I$ (and possibly $S_A$), and misbehaving parties are slashed. If the claim is invalid, the escalator loses its bond and pays verification fees.
\end{enumerate}

\begin{figure}[ht]
    \centering
    \includegraphics[width=1\linewidth]{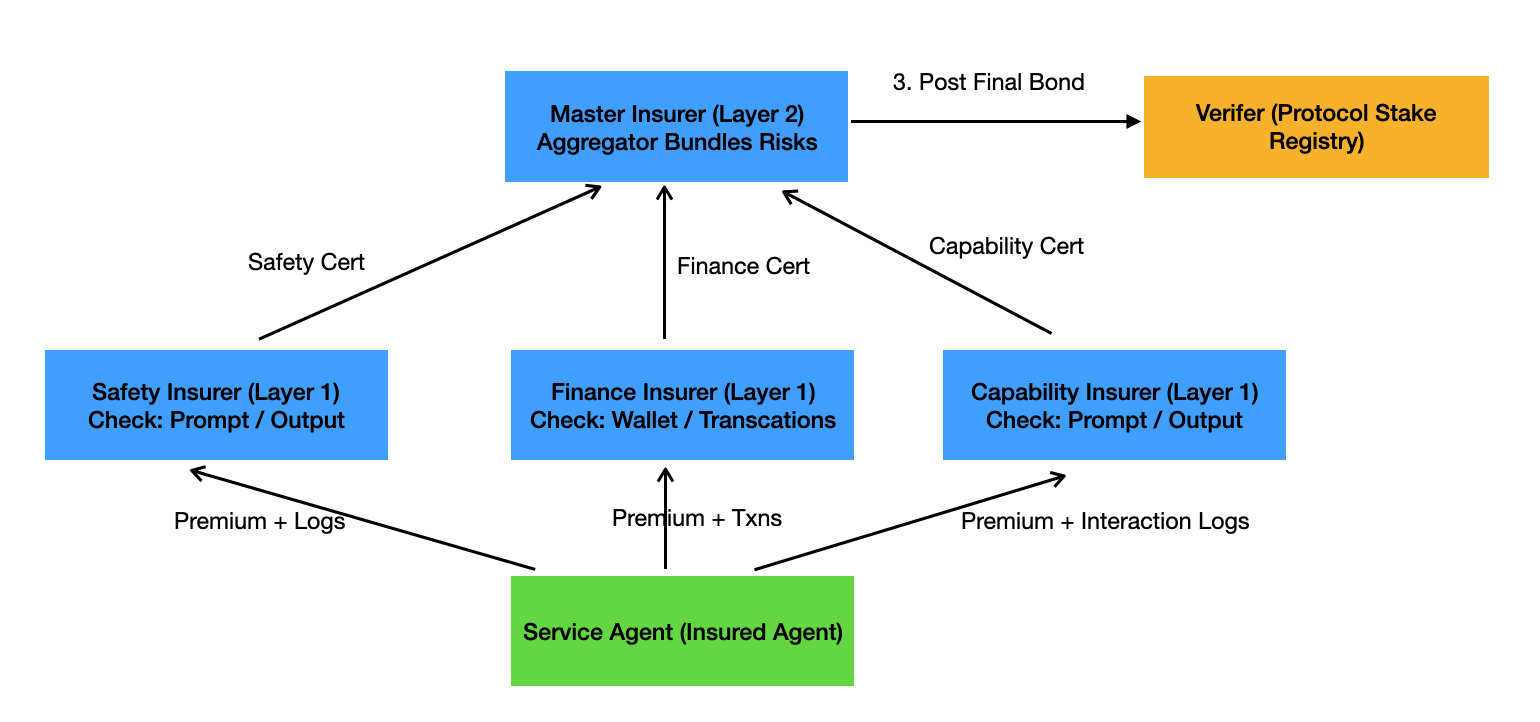}
    \caption{The hierarchical insurance structure}
    \label{fig:stack}
\end{figure}

\subsection{The Economic Feedback Loop}

In the insured-agent pattern, economic interactions form a simple loop. First, risk is transferred: the insurer posts the required collateral (stake) to the protocol registry on the agent’s behalf, and if the agent is slashed, the insurer—not the agent—loses this capital. Second, the agent pays recurring premiums to the insurer, with the fee dynamically priced according to the agent’s codebase, behavioral history, and implemented safeguards. Third, liability naturally rests with the insurer, which aligns incentives: because their own capital is at risk, insurers are financially motivated to rigorously vet agents before onboarding them and to continuously monitor their behavior during operation.

\subsection{Privacy-preserving verification}

A central challenge for verification is privacy: agents cannot simply publish all internal logs on-chain for verification without leaking sensitive data. The insurance pattern addresses this via privileged audit access. The agent needs insurance to participate in high-stakes interactions demanded by clients, while the insurer needs sufficient visibility into the agent’s behavior to avoid large losses. As a result, the agent voluntarily grants the insurer special access—such as read permissions to private execution logs or remote attestation keys for a Trusted Execution Environment (TEE). Verification thus becomes a mutually desired event between agent and insurer: no global authority coerces the agent to reveal data publicly; instead, the agent selectively discloses information to a chosen insurer in exchange for coverage and capital. This creates a voluntary verification market rather than a centralized surveillance regime.

\subsection{The hierarchical insurance structure}

Verification needs differ across domains, so a one-size-fits-all approach is inadequate. We therefore propose a hierarchical insurance structure in which multiple specialized insurers collaborate to underwrite a single agent (see Fig. \ref{fig:stack}). In this stack, different insurers occupy distinct layers and bring domain-specific verification expertise. For example, a Safety Insurer at Layer 1 focuses on large language models: it audits prompt design, tests jailbreak robustness, and monitors semantic logs. A Financial Insurer, also at Layer 1, specializes in DeFi and smart contracts: it validates the agent’s wallet constraints (e.g., “never transfer more than 1 ETH”) and monitors transaction traces. A Master Insurer at Layer 2 then aggregates these lower-layer policies. If an agent holds both a “Safety Certificate” and a “Financial Certificate,” the Master Insurer can treat the residual risk as lower and is willing to post the final protocol bond. This yields a composable trust ecosystem: new agent developers need not accumulate a massive standalone reputation; instead, they can pass audits by respected specialist insurers. The trust attached to an agent is effectively the aggregate reputation and certified assurances of the insurers that stand behind it.

\subsection{Game-Theoretic Model}

We model a single insured interaction as a sequential game with three strategic players $A$, $I$, $U$ and a deterministic oracle $V$. The parameters are: 
\begin{description}
    \item[$L$] Loss to $U$ if $A$ misbehaves.
    \item[$G$] One-shot gain to $A$ from misbehaving.
    \item[$S_A$] Agent stake (deductible) locked with $I$.
    \item[$S_I$] Insurer stake locked in the protocol.
    \item[$B$] Escalation bond posted by both $U$ and $I$ in a dispute.
    \item[$F$] Fee paid to $V$ when called.
    \item[$R$] Reputation cost to $I$ if caught denying a valid claim.
    \item[$V_{\text{future}}$] Discounted value of $A$’s future business and insurability.
\end{description}

\noindent
\textbf{Assumptions.}
\begin{itemize}
    \item Players are rational, risk-neutral, and share common knowledge of the mechanism.
    \item If invoked, $V$ reveals whether $A$ acted honestly ($H$) or maliciously ($M$) without error.
    \item Solvency: $S_I \geq L$ for the insured range of losses.
\end{itemize}

The extensive-form structure is:
\begin{enumerate}
    \item \textbf{Stage 1 (Agent).} $A$ chooses $H$ or $M$. Under $M$, $A$ gains $G$ and $U$ suffers loss $L$.
    \item \textbf{Stage 2 (User).} Observing $L$, $U$ chooses to file a claim $C$ or not $NC$.
    \item \textbf{Stage 3 (Insurer).} On $C$, $I$ inspects logs and chooses to accept (Acc) or deny (Den).
    \item \textbf{Stage 4 (Escalation).} If Den, $U$ chooses to escalate (Esc) or drop (Drop). On Esc, both post bond $B$, $V$ is called, and the losing side forfeits its bond and pays $F$.
\end{enumerate}

\subsection{Equilibrium and Proof}

\begin{theorem}
If the following constraints hold:
\begin{align}
    &\text{(Access to justice)} && 2L + B > F, \label{eq:justice}\\
    &\text{(Solvency)} && S_I \geq L, \label{eq:solvency}\\
    &\text{(Deterrence)} && S_A + V_{\text{future}} > G, \label{eq:deterrence}
\end{align}
then there exists a subgame-perfect equilibrium in which $A$ acts honestly, $I$ pays all valid claims and rejects invalid ones, and $U$ escalates only valid disputes.
\end{theorem}

\begin{proof}
We proceed by backward induction. 

\emph{Stage 4 (User).} For a valid claim (agent misbehaved), escalation yields
\[
\pi_U^{\text{Esc}} = L + B - F,
\]
while dropping yields $\pi_U^{\text{Drop}} = -L$. Condition~\eqref{eq:justice} implies $\pi_U^{\text{Esc}} > \pi_U^{\text{Drop}}$, so $U$ escalates any valid claim. For an invalid claim, escalation gives $\pi_U^{\text{Esc,inv}} = -B - F < 0$ whereas dropping gives $0$, so $U$ never escalates invalid claims.

\emph{Stage 3 (Insurer).} For a valid claim that will be escalated if denied, $I$ compares
\[
\pi_I^{\text{Acc}} = -L + S_A,\qquad
\pi_I^{\text{Den}} = -L - B - F - R.
\]
With $S_A, B, F, R > 0$, we have $\pi_I^{\text{Acc}} > \pi_I^{\text{Den}}$, so $I$ accepts all valid claims. For invalid claims, paying $L$ is strictly worse than denying and winning escalation, so $I$ rejects invalid claims.

\emph{Stage 1 (Agent).} Anticipating that valid claims are paid and that misbehavior is detectable, a malicious deviation yields
\[
\pi_A^{M} = G - S_A - V_{\text{future}},
\]
while honest behavior yields $\pi_A^{H} = \Pi_{\text{honest}}$ (which includes preserving stake and future business). Condition~\eqref{eq:deterrence} implies $\pi_A^{M} < \Pi_{\text{honest}}$ for any reasonable $\Pi_{\text{honest}}$, so honesty strictly dominates misbehavior.

Finally, \eqref{eq:solvency} guarantees that $I$ can always pay $L$ when required, so promised payouts are credible. Hence, the strategy profile ``$A$ honest; $U$ claims when harmed and escalates only valid denials; $I$ pays valid claims and rejects invalid ones'' constitutes a subgame-perfect equilibrium.
\end{proof}

Economically, the insured-agent mechanism creates an ``optimistic'' structure: the costly verifier $V$ is rarely invoked in equilibrium, yet its availability and the threat of slashing ensure that insurers and agents behave honestly in almost all on-chain interactions.

\section{Discussion: Open Challenges and Research Agenda}

We close by outlining a concise research agenda at the intersection of MAS, cryptoeconomics, and HCI opened up by insured agents.

\paragraph{Underwriting and risk calibration as MAS mechanism design.}
Underwriting becomes the problem of translating heterogeneous safety evidence (red-teaming, jailbreak robustness, sandboxed tool-use tests) into premiums and collateral for diverse tasks under distribution shift and strategic behavior \cite{Rothschild1976Equilibrium}. Insurers need incentive-compatible risk metrics, belief-updating rules that protect sensitive model/prompt information, and portfolio models for correlated failures induced by shared base models and common attack surfaces such as hallucination, and prompt injection.

\paragraph{Moral hazard, adverse selection, and “insured misbehavior.”}
Insurance reshapes incentives and introduces classic problems of moral hazard and adverse selection: high-risk agents may seek coverage disproportionately, and insured agents may take greater risks \cite{Rothschild1976Equilibrium}. In agentic settings, these are exacerbated by rapid model updates, hidden prompt changes, and toolchain drift. Classic instruments—deductibles, exclusions, experience rating—need MAS-specific forms \cite{Pinyol2013Computationala}. Decentralized markets must also address insurer–agent collusion, sybil strategies, and platform capture, calling for insurer reputation, challenge markets, and reinsurance designs that reduce systemic fragility without enabling cartels \cite{Josang2007survey}.

\paragraph{Evidence design, verifiable misbehavior, and proof-carrying interactions.}
Dispute resolution requires operational definitions of “misbehavior” and standardized, privacy-respecting evidence formats. For LLM agents, misbehavior includes violating constraints, leaking secrets, or acting outside mandate; for tool-using agents, unauthorized calls or unsafe transactions. Expressive but enforceable policy languages (e.g., “max drawdown,” “never transfer $> X$,” “only call tool $T$ with inputs matching schema”) are a MAS protocol challenge \cite{Mohammad2025}. Insured agents motivate tamper-evident logs, signed traces, and secure replay that bind actions to principals without a single central logger. At scale, this suggests “proof-carrying interactions,” \cite{Necula1997Proofcarrying} where agents attach succinct cryptographic evidence of key properties using practical secure multi-party computation (MPC) \cite{Zhao2019Secure} and zero-knowledge tools (ZK) \cite{DeSantis1992Zeroknowledge}. 

\paragraph{Privacy, auditability, and the limits of TEEs.}
Auditability introduces a privacy–accountability tradeoff. TEEs and trusted hardware can support remote attestation and reduce leakage, but remain vulnerable to side channels and implementation flaws \cite{Munoz2023surveya}. A defense-in-depth agenda combines TEEs with redundancy and cryptographic commitments, plus “least-privilege audits” that reveal only what is necessary for claims. Differentially private audit summaries and hybrid on/off-chain attestations are promising but underexplored for agentic systems \cite{Dwork2006Differential}.

\paragraph{Governance, contestability, and human-facing interfaces.}
Insured agents reframe trust as a process of coverage, claims, and contestation. Users need intelligible coverage terms, clear escalation paths, and usable evidence interfaces, aligning with work on contestability and procedural fairness in automated decisions \cite{dignum2025contesting, kroll2020accountability}. HCI challenges include communicating risk and exclusions, avoiding over-trust in “insurance-backed” agents, and enabling meaningful recourse without excessive friction. Prior work on trust calibration and human–AI decision support is directly relevant \cite{Amershi2019Guidelines}. Designing how insurance signals appear in explainable AI and UIs (e.g., badges, dashboards, claim histories) is itself a key part of the agenda.

\paragraph{Interoperability and standardization in the agentic web.}
For adoption, insurance primitives must integrate with emerging agentic-web stacks. This entails standardizing policy metadata, insurer attestations, and audit-log schemas so that discovery registries and payment rails can consume insurance signals. ERC-8004 already anticipates "agent-at-staking" mechanism; insured agents provide a concrete design space for how these markets operate and which hooks they require \cite{erc8004}. Open questions include representing and transporting “proof of insurance” across A2A-style interactions and payment protocols such as AP2, and ensuring standards remain neutral to particular insurer or verifier architectures.

\paragraph{Evaluation methodology for insured agents.}
An effective insurance proposal requires measurable impact. Insured agents call for benchmarks that track not only task success, but also: (i) economic loss distributions under adversarial conditions, (ii) deterrence effects—how misbehavior rates change with insurance, (iii) dispute frequency and resolution time, (iv) privacy leakage during audits, and (v) market health indicators such as competition, entry barriers, and signs of cartelization.

\section{Conclusion}
Open agent economies require credible trust without assuming perfectly reliable agents or centralized verifiers. We propose insured agents as a protocol-native institution for the agentic web: competitive insurers post slashable stake on behalf of agents, price risk through premiums, and provide privacy-preserving verification through privileged audit access. A hierarchical insurer market supports specialization, while an optimistic escalation game sustains incentive compatibility with rare recourse to expensive verifiers.

\bibliographystyle{ACM-Reference-Format}
\bibliography{references}

\end{document}